\documentstyle[preprint,aps,floats]{revtex}
\begin{document}
\draft
\preprint{} 
\title{Exact Solution of Frenkel-Kontorova Models with a Complete Devil's 
Staircase in Higher Dimensions}
\author{Hsien-chung Kao$^{1,2}$, Shih-Chang Lee$^1$, and Wen-Jer Tzeng$^1$}
\address{$^1$Institute of Physics, Academia Sinica, Taipei, Taiwan 11529, R.O.C.}
\address{$^2$Department of Physics, Tamkang University, Tamsui, Taiwan 25137, 
R.O.C.}

\maketitle
\begin{abstract}
We solve exactly a class of Frenkel-Kontorova models with 
piecewise parabolic potential, which has $d$ sub-wells in a period.  With 
careful analysis, we show that the phase diagram of the minimum enthalpy 
configurations exhibits the structure of a complete $d$-dimensional devil's 
staircase.  The winding number of a minimum enthalpy configuration is locked 
to rational values, while the fraction of atoms in each sub-well is locked 
to values which are sub-commensurable with the winding number.
\end{abstract}
\narrowtext

Periodic modulated structures are quite common in both condensed-matter and 
dynamical systems.  As suitable parameters are varied, such structures may
go through commensurate-incommensurate phase transitions.  In particular, when 
a system is in the highly nonlinear regime, there is generally a tendency for 
the periodicity to lock into values which are commensurable with the lattice 
constant\cite{bak82}.  The Frenkel-Kontorova (FK) model is just such an 
example, which describes a one-dimensional chain of coupled atoms subject to a
periodic potential $V$: 
\begin{equation}
H(\{u_n\}) =
\sum_n \left[\frac{1}{2}\left(u_{n+1}-u_n \right)^2+\lambda V(u_n) 
- \sigma \left(u_{n+1}-u_n\right) \right].
\label{eq:ham}
\end{equation}
Here, $u_n$ is the position of the $n$th atom, $\lambda $ is a measure of the 
nonlinearity.  For simplicity, suitable length
scale is chosen so that the periodicity of $V(u)$ is one.

From the variational principle, we see that a stationary configuration 
$\{u_n\}$ must satisfy the following equation:
\begin{equation}
u_{n+1}-2u_n+u_{n-1}=\lambda V'(u_n).
\label{eq:forcebal}
\end{equation} 
It is well-known that the above equation can be rewritten as coupled first 
order difference equations, which provides a link between the stationary 
configurations in an FK model and the trajectories in a two-dimensional 
area-preserving map\cite{aub78}.

According to Aubry\cite{aub80i,aub83}, a minimum-energy configuration 
is one in which $H$ cannot be decreased by alteration of a finite 
number of $u_n$'s.  For these configurations, there is a well-defined winding 
number 
\begin{equation}
\omega \equiv \lim_{N,N'\to \infty}\frac{u_N-u_{-N'}}{N+N'},
\label{eq:wind}
\end{equation}
which is the mean distance between neighboring atoms.  A ground state is a 
recurrent minimum-energy configuration, which can be depicted
by\cite{aub80i,aub83}
\begin{equation}
u_n=f_\omega(n\omega+\alpha).
\label{eq:hulldef}
\end{equation}
Here, $\alpha $ is an arbitrary  phase variable; the hull function 
$f_\omega(x)$ is increasing with $x$ and satisfies
\begin{equation}
f_\omega(x+1)=f_\omega(x)+1.
\end{equation}
When $f_\omega(x)$ is discontinuous, the positions of 
the atoms is described by either $f_\omega(x^+)$ or $f_\omega(x^-)$.

There are two distinct types of phase transitions in the FK models with
smooth potential.  First by varying $\sigma$ with $\lambda $ fixed, one 
finds a commensurate-incommensurate phase transition.  For a given $\sigma$, 
there is a $\omega(\sigma)$ which specifies the minimum enthalpy state and it 
takes on rational value on an infinite set of plateaus.  Such structures are 
called incomplete or complete devil's staircases depending on whether or not
there are incommensurate phases with non-zero measure between the 
commensurate ones.  \cite{bak82,aub78,aub80i,aub83,aub80,bak86}.  There exists
a critical value of $\lambda$, above which the devil's staircase becomes 
complete.  On the other hand, if $\omega $ is fixed while $\lambda $ is 
varied, one will see that for the incommensurate ground states  
there is a non-vanishing Peierls-Nabarro barrier above a critical value of 
$\lambda$\cite{aub78,KLT}.  It is called the transition by breaking of 
analyticity (TBA), since the most evident indication of this transition is 
the change in the analyticity of the hull function\cite{aub78}.  In terms of 
the two-dimensional area-preserving map, what happens to an incommensurate 
ground state is that it changes from a configuration represented by a KAM 
torus to one which is represented by a Cantor-Aubry-Mather (CAM) 
set\cite{aub91}.  

The first exactly solved FK model was studied by Aubry using a piecewise 
parabolic potential \cite{aub78,aub80,aub-dev}.  The potential is the 
simplest non-linear one in that the non-linearity only occurs at one point in 
a period.  It turns out that in this model both the critical values of 
$\lambda$ are zero for the transitions described above.  In other words, 
the devil's staircase in this system is always complete and the incommensurate
configuration is always pinned unless $\lambda=0$.  Simple as the model may 
seem, it provides us a clear picture of systems in the pinned phase.

In this paper, we will solve exactly an extension of the Aubry model, 
when the potential has $d$ sub-wells in a period.  Such substructure may 
occur naturally in surface reconstruction or be realized in a superlattice.  
This model was first proposed by Griffiths, et al \cite{Griffiths}.  Several 
interesting new phenomena such as solitons in the incommensurate phase,
discontinuous Cantorus-Cantorus phase transitions, etc were found in the $d=2$
case.  As we shall see, the phase diagrams in the higher dimensional case are
much more complicated

When there are $d$ pieces of parabolas in a period, the potential is given by
\begin{equation}
V(u)= \min_{1\le i \le d} \bigl\{V_i(u)\bigr\},
\label{eq:v(u)}
\end{equation} 
with 
\begin{equation}
V_i(u) = \frac{1}{2}\left(u - b_i - {\rm Int}[u]\right)^2 + h_i.
\label{eq:vi(u)}
\end{equation} 
Here, ${\rm Int}[u] $ is the integer part of $u$; $h_i$ and $b_i$'s are generally 
independent parameters.  Requiring $V(u)$ to be continuous, we find that
the positions of the kinks are given by
\begin{equation}
p_i = \frac{b_i + b_{i+1}}{2} + \frac{h_{i+1}- h_i}{b_{i+1} - b_i},
\end{equation}
with $h_{d+i} = 1+ h_i$ and $b_{d+i} = 1+ b_i$. 
$p_i$'s must satisfy
\begin{equation}
0 \le p_1 \le \ldots \le p_{d} \le 1.
\label{eq:alpha}
\end{equation}
For convenience, we will choose $p_d =1$, which leads to the constraint 
$h_1 + b_1^2/2 = h_d + (1-b_d)^2/2 $.  An example for $d = 3$ is shown in 
Fig. 1.

The force-balance equation is given by
\begin{equation}
u_{n+1} + u_{n-1} - (2+\lambda)u_n = -\lambda \left({\rm Int}[u_n] + b_i \right), 
\qquad p_{i-1} \le u_n - {\rm Int}[u_n] \le p_i.
\label{eq:eom}
\end{equation} 
Introducing the hull function $f_\omega(x)$\cite{fn0} and making use of 
the fact that it is increasing with $x$\cite{KLT}, we can rewrite the above 
equation as
\begin{equation}
f_\omega(x+\omega) + f_\omega(x-\omega) 
- (2 + \lambda)f_\omega(x) = -\lambda
\left\{b_1 + \sum_{i=1}^{d} (b_{i+1} - b_i) {\rm Int}[x+1- \beta_i] \right\},
\label{eq:eomf}
\end{equation}
where $\beta_i \equiv \sum_{j=1}^{i} \nu_j$.  
$\nu_i$ designates the fraction of atoms that lies in the $i$-th 
sub-well and $\sum_{i=1}^{d} \nu_i = 1.$  For consistency, 
$\beta_i$ must satisfy the following condition:
\begin{equation}
f_\omega(\beta_i^-) \le p_i \le f_\omega(\beta_i^+).
\label{eq:consistency}
\end{equation}
It is easy to verify that the solution to Eq.~(\ref{eq:eomf}) is given by
\begin{eqnarray}
f_\omega(x) 
= d_0 \sum_{n= -\infty}^{\infty} {\rm e}^{-|n|\chi} 
\left\{b_1 + \sum_{i=1}^{d} (b_{i+1} - b_i){\rm Int}[n\omega+x+1-\beta_i] 
\right\}.
\label{eq:hull}
\end{eqnarray} 
Here ${\rm e}^{-\chi} \equiv 
\left\{1 + \lambda/2 - \sqrt{\lambda(1+\lambda/4)} \;\right\}$, and 
$d_0 \equiv \tanh(\chi/2).$  It is obvious from the above result that around the
$i$-th kink at $p_i$ there will be a region of length $(b_{i+1} - b_i)d_0$, 
in which no atom is allowed.

Once we know the hull function, it is straightforward to find the average 
enthalpy per atom:
\begin{eqnarray}
\eta(\omega, \nu_1, \ldots, \nu_d) 
& = &\frac{\left(\omega^2 -2\sigma \omega\right)}{2} 
+ \frac{\lambda}{2}\sum_{1\le i<j \le d} 
\biggl\{ (b_i - b_j)(1 + b_i - b_j)\nu_i\nu_j \biggr\}
+ \lambda \sum_{i=1}^{d} h_i \nu_i
\nonumber \\
& \; & - \frac{\lambda\, d_0}{4}\sum_{n= -\infty}^{\infty} 
{\rm e}^{-|n|\chi} \left\{ 
\sum_{i,j = 1}^{d} (b_{i+1} - b_i)(b_{j+1} - b_j) 
\,S(n\omega + \beta_{j,i}) \right\}.
\label{eq:enthalpy}
\end{eqnarray}
Here, $S(x) \equiv \left(x - {\rm Int}[x] \right)^2 -\left(x - {\rm Int}[x] \right),$
and $\beta_{j,i} \equiv \beta_j - \beta_i$.  The hull function and the average
energy per atom have been obtained independently by Griffiths, et al in a 
different form\cite{note}.

Since $\eta(\omega, \nu_1, \ldots, \nu_d)$ is a convex non-differentiable 
function, the minimum enthalpy configuration for a given 
set of \mbox{$(\sigma, h_1, \ldots, h_d, b_1, \ldots, b_d)$} must be
determined by the condition that the directional derivatives are non-negative
in all directions.  After some work, the condition simplifies to
\begin{eqnarray}
& \cos\theta_0 &\left\{-\sigma 
+ \frac{\lambda\, d_0}{2} \sum_{n= -\infty}^{\infty} n\, {\rm e}^{-|n|\chi}
\biggl( \sum_{j,k =0}^{d-1}\Delta b_j \Delta b_k
{\rm Int}[n\omega + \beta_{k,j} + \delta^n_{k,j} ] \biggr)\right\} \nonumber \\
& + \lambda\;\sum\limits_{i=1}^{d-1} \cos\theta_i &  
\left\{ \Delta h_i + d_0 \sum_{n= -\infty}^{\infty} {\rm e}^{-|n|\chi}
\biggl(\sum_{j=0}^{i-1} \sum_{k=i}^{d-1} \Delta b_j \Delta b_k
{\rm Int}[n\omega + \beta_{k,j} + \delta^n_{k,j}] \biggr)\right\} \ge 0. 
\label{eq:GSd} 
\end{eqnarray}
Here, $\Delta h_i \equiv (h_{i} - h_d)$,
$\Delta b_j \equiv (b_{j+1} - b_j)$,
$\delta^n_{k,j} \equiv \epsilon(n\cos\theta_0 
+\;\sum\limits_{l= 1}^{k}\cos\theta_l -\;\sum\limits_{l= 1}^{j}\cos\theta_l)$.
$\cos\theta_i$'s satisfy $\sum\limits_{i=0}^{d-1} \cos^2\theta_i =1,$ 
and $0 \le \theta_i \le \pi.$  $\delta^n_{k,j}$ is just a bookkeeping device to
remind one of which limit to take in the function ${\rm Int}[x]$ at integer values 
of $x$.  It is obvious 
from the above result that the phase diagram only depends on 
$(\Delta h_1, \ldots, \Delta h_{d-1}, \Delta b_1, \ldots, \Delta b_{d-1}),$ 
as it should be.

To illustrate how the phase diagrams look like, let's first concentrate on the 
case of $d=2.$  For a given set of $(\omega, \Delta h_1, \Delta b_1)$, 
there is only one $\nu_1$ that gives rise to the minimum energy state, and it 
can be determined by setting $\theta_0 = \pm \pi/2$ in the minimum enthalpy
condition, Eq.~(\ref{eq:GSd}).  The resulting $\nu_1$ is locked to values
sub-commensurate to $\omega$, i.e. there exists some integer $n$ such that 
$n\omega + \nu_1 \in {\bf Z}$ (the set of all integers).  
In Fig.~2, we show the phase diagram for $\omega = (\sqrt{5} -1)/2.$  
Each vertical cross section will give a devil's staircase in the 
$\Delta h_1$-$\nu_1$ plane.  A special case for $\Delta b_1 =1/2$ has been 
shown in Fig.~2 of \cite{Griffiths}.

The phase diagram of the minimum enthalpy states in this case is three 
dimensional.  For simplicity, let's look at the constant $\Delta b_1 $ and 
$\Delta h_1$ slices, respectively.  In the former case, the domains of 
stability fall into three categories (see Fig.~3):

\begin{description}
\item{i.)}\,$n_0\omega, (n_1\omega + \nu_1) \not\in {\bf Z}$, 
for any $n_0, n_1 \in {\bf Z}$ ($\omega $ irrational): 

The domains are discrete points. 

\item{ii.)}\, There exists $n \in {\bf Z}$ such that either $n\omega $ or
$(n\omega + \nu_1) \in {\bf Z}.$
\begin{enumerate}
\item \,$\omega = p/q$ \cite{fn1}.  $(n\omega +\nu_1) \not\in {\bf Z}$ for 
any $n \in {\bf Z}$: 

The domains are horizontal lines with length 
$\frac{\textstyle (2\Delta b_1^2 - 2\Delta b_1 + 1) \lambda\, d_0\, q\, 
{\rm e}^{-q\chi}} {\textstyle (1-{\rm e}^{-q\chi})^2}.$

\item \,$\omega $ irrational.  $(n\omega + \nu_1) \in {\bf Z}$ for some integer 
$n:$

The domains are lines with slope $-1/(\lambda n)$ and length 
\mbox{$\Delta b_1 (1 - \Delta b_1) \sqrt{1+ \lambda^2 n^2} 
\left(d_0 {\rm e}^{-|n|\chi}\right).$}
\end{enumerate}

\item{iii.)}\, $\omega = p/q,$ $(s\omega + \nu_1) \in {\bf Z}$ for 
some integer $s \in [0, q-1]:$

The domains are infinite-sided polygons with sides given by the lines 
described in (ii) and vertices by the points in (i).  
\end{description}

Summing the area of all these polygons, we verify that they form a 
two-dimensional analog of a complete devil's staircase in the
$\sigma-\Delta h_1 $ plane.  The phase diagram we obtained here is very 
similar to that obtained by Aubry, et al in a different model \cite{dev2} 
and that obtained numerically by P. Delaly\cite{thesis}.  

For constant $\Delta h_1$, the phase diagram is very similar to the previous 
one, except for the following two aspects (see Fig.~4):
\begin{description}
\item{a.)}\, One can easily see from Fig.~2 that only some of the $\nu_1$'s 
satisfying the sub-commensurability condition are allowed; 

\item{b.)}\, the lines in (ii.2) are parabolas instead.
\end{description}

It is straightforward to carry out a similar analysis for the phase diagrams
in higher dimensions, i.e. $d \ge 3$.  Let us introduce 
$\sigma = \overline{\sigma} +\delta\sigma$ and 
$\Delta h_i = \overline{\Delta h_i} +\delta h_i$. 
Here,
\begin{eqnarray}
& \overline{\sigma} & \quad = \;
\frac{\lambda\, d_0}{2} \sum_{n= -\infty}^{\infty} n\, {\rm e}^{-|n|\chi}
\biggl\{ \sum_{j,k =0}^{d-1}\Delta b_j \Delta b_k
\overline{{\rm Int}}[n\omega + \beta_{k,j} ] \biggr\}, \nonumber \\
& \overline{\Delta h_i} & \quad = \;
d_0 \sum_{n= -\infty}^{\infty} {\rm e}^{-|n|\chi}
\biggl\{\sum_{j=0}^{i-1} \sum_{k=i}^{d-1} \Delta b_j \Delta b_k
\overline{{\rm Int}}[n\omega + \beta_{k,j} ] \biggr\}, 
\label{eq:center} 
\end{eqnarray}
with $\overline{{\rm Int}}[x] \equiv ({\rm Int}[x^+]+{\rm Int}[x^-])/2.$
Eq.~(\ref{eq:GSd}) becomes
\begin{eqnarray}
\hat{\bf u}\cdot {\bf v}  
&\le & \frac{\lambda\, d_0}{4} \sum_{n= -\infty}^{\infty} {\rm e}^{-|n|\chi}
\biggl\{ \sum_{j,k =0}^{d-1}\Delta b_j \Delta b_k Z(n\omega + \beta_{k,j} ) 
|{\bf V}^n_{k,j}\cdot \hat{\bf u}|\biggr\} .
\label{eq:region} 
\end{eqnarray}
Here, $Z(x)$ takes the value 1 or 0 depending on whether $x$ is an integer or 
not. \break 
\mbox{
$\hat{\bf u} \equiv (\cos\theta_0, \cos\theta_1, \ldots, \cos\theta_{d-1})$,}
an arbitrary unit vector, and 
\mbox{
${\bf v} \equiv 
(\delta\sigma, - \lambda \delta h_1, \ldots,- \lambda \delta h_{d-1})$.}
\mbox{
${\bf V}^n_{k,j} \equiv (n, 0, \ldots, 0, 1, \ldots, 1, 0, \ldots, 0)$,}
where the value 1 is assigned to the $[\min(j,k)+2]$-th to $[\max(j,k)+1]$-th 
components.
It can be seen that for a domain of stability to have non-zero measure, 
the corresponding $\omega$ must be rational and all the $\nu_i$'s must
be sub-commensurate to $\omega$.  Furthermore, these domains are enclosed by 
infinitely many hyper-planes, where either $\omega$ is irrational while all 
the sub-commensurability conditions are satisfied or $\omega$ is rational 
while exactly one of the sub-commensurability conditions is violated.  
The intersections of these hyper-planes in turn give the boundaries of even 
lower dimensions, where two or more of the commensurability and 
sub-commensurability conditions fail.  They correspond to multi-critical points
in the phase diagram.  

In this letter, we solve exactly a class of extended solvable FK models,  
whose potential has $d$ sub-well in a period.  Both the hull function and the 
enthalpy per atom for an arbitrary $d$ are given explicitly.  Since the 
enthalpy per atom is non-analytic, the global minimum is found 
by requiring all the directional derivatives to be non-negative.  Using this 
condition, we are able to determine analytically the complete phase diagram.
For given $\Delta b_i$'s, the phase diagram is an extension of the devil's 
staircase to $d$ dimensions as long as $\lambda > 0$.  It will be interesting
to see if the conclusion is still valid in the $d \to \infty$ limit.
Detailed discussion of the phase diagram and the associated physics will be 
given in a subsequent paper.

\acknowledgments
This work is supported in part by grants from the National Science
Council of Taiwan-Republic of China under the contract number
No. NSC85-2112-M-001-004.

\vfill\eject
    
\begin{figure}[h]
\includegraphics{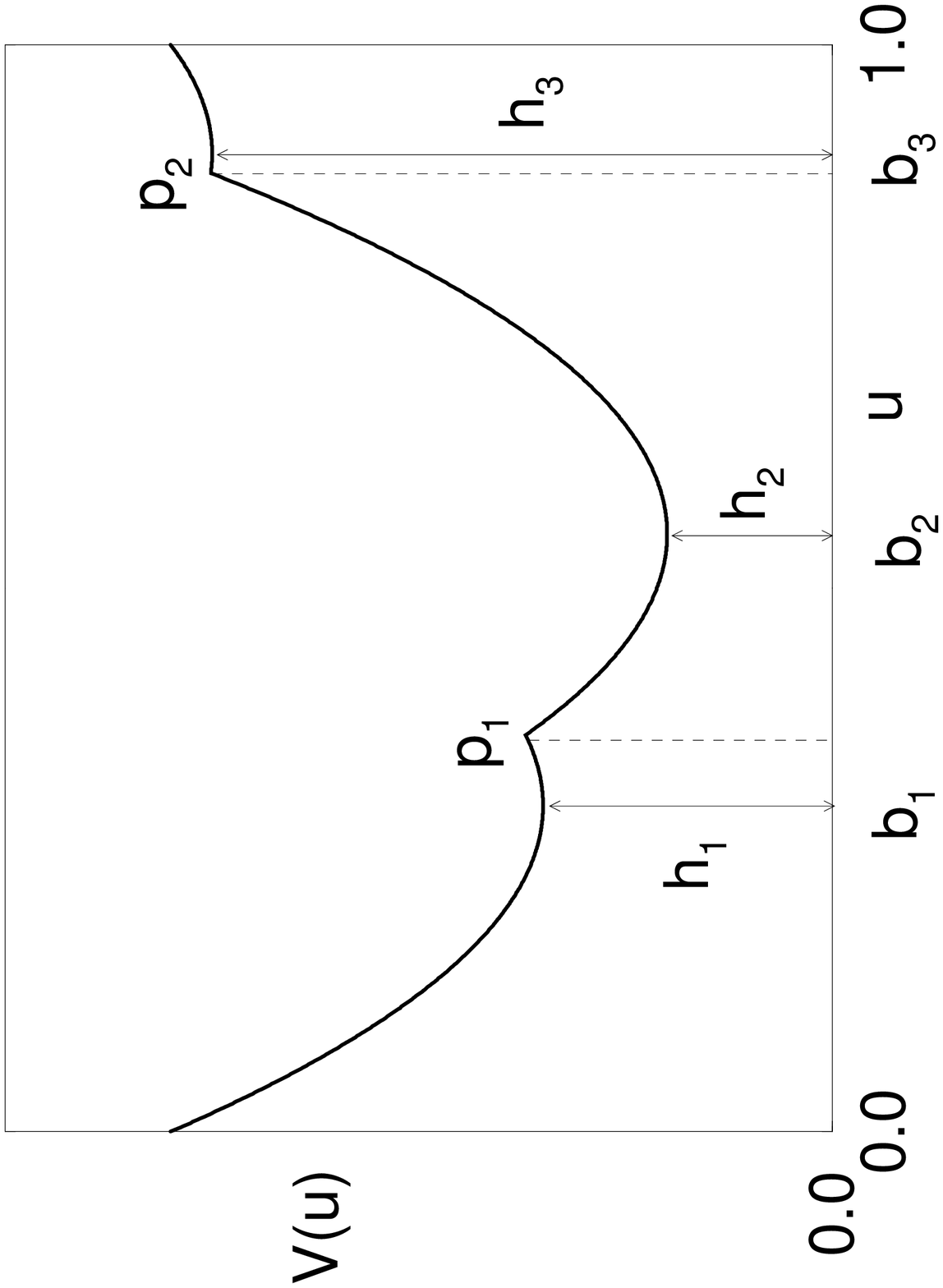}
\vskip 12.5cm
\end{figure}
$\qquad\qquad\qquad\qquad $ 
Fig.~1~~  Periodic potential $V$ in Eq.(6) for d=3.
\vfill\eject

\begin{figure}[h]
\includegraphics{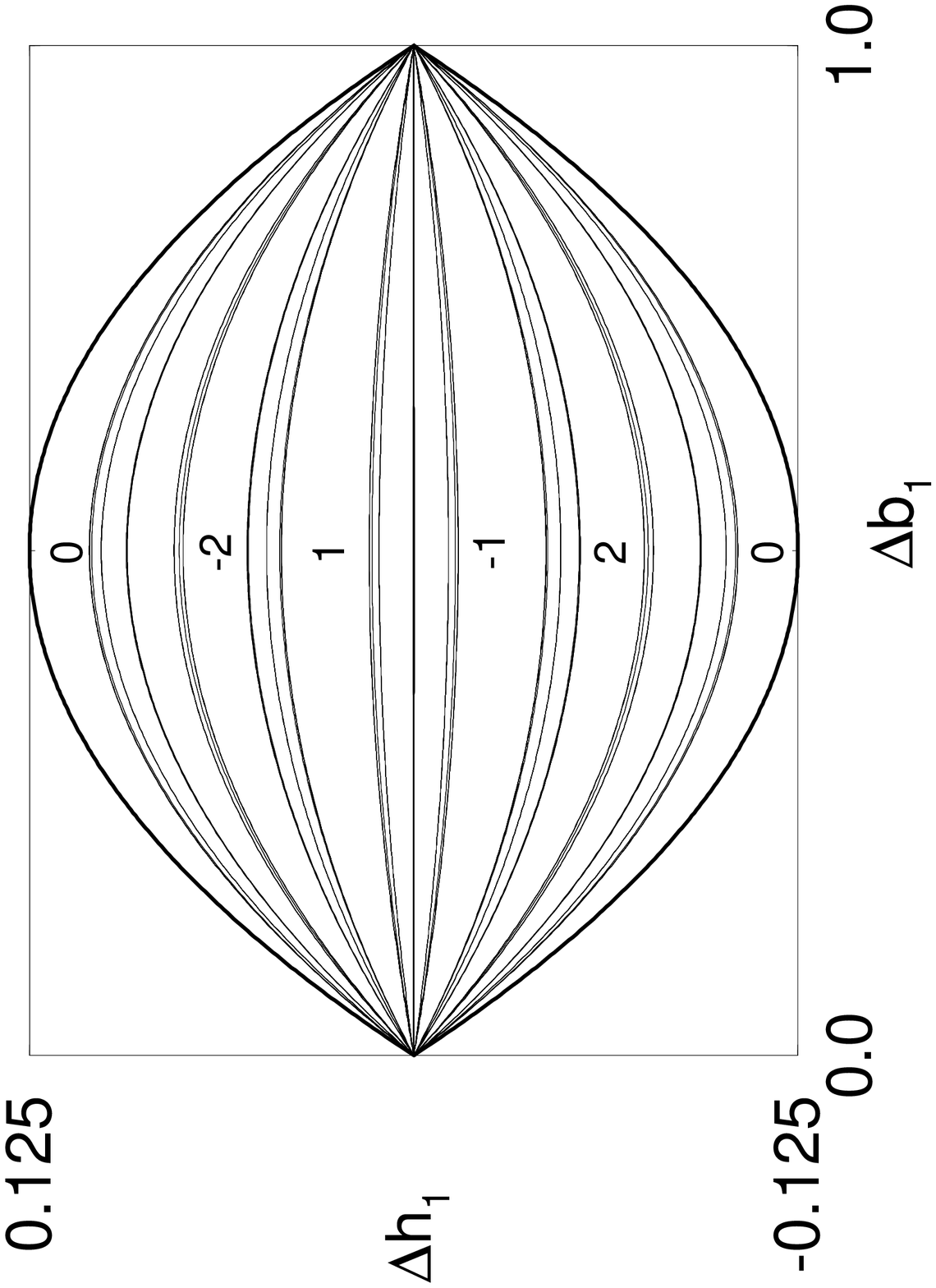}
\vskip 12.5cm
\end{figure}
\noindent
Fig.~2~~  Phase diagram in the $\Delta b_1$-$\Delta h_1$ plane for 
 $\lambda =0.1$, and $\omega = (\sqrt{5}-1)/2$.  The values of $n$ are inset,
which is related to $\nu_1$ through $\nu_1 = 1+ {\rm Int}[n\omega]- n\omega $.
The two heavy lines give the upper and lower bound of $\Delta h_1$.
\vfill\eject

\begin{figure}[h]
\includegraphics{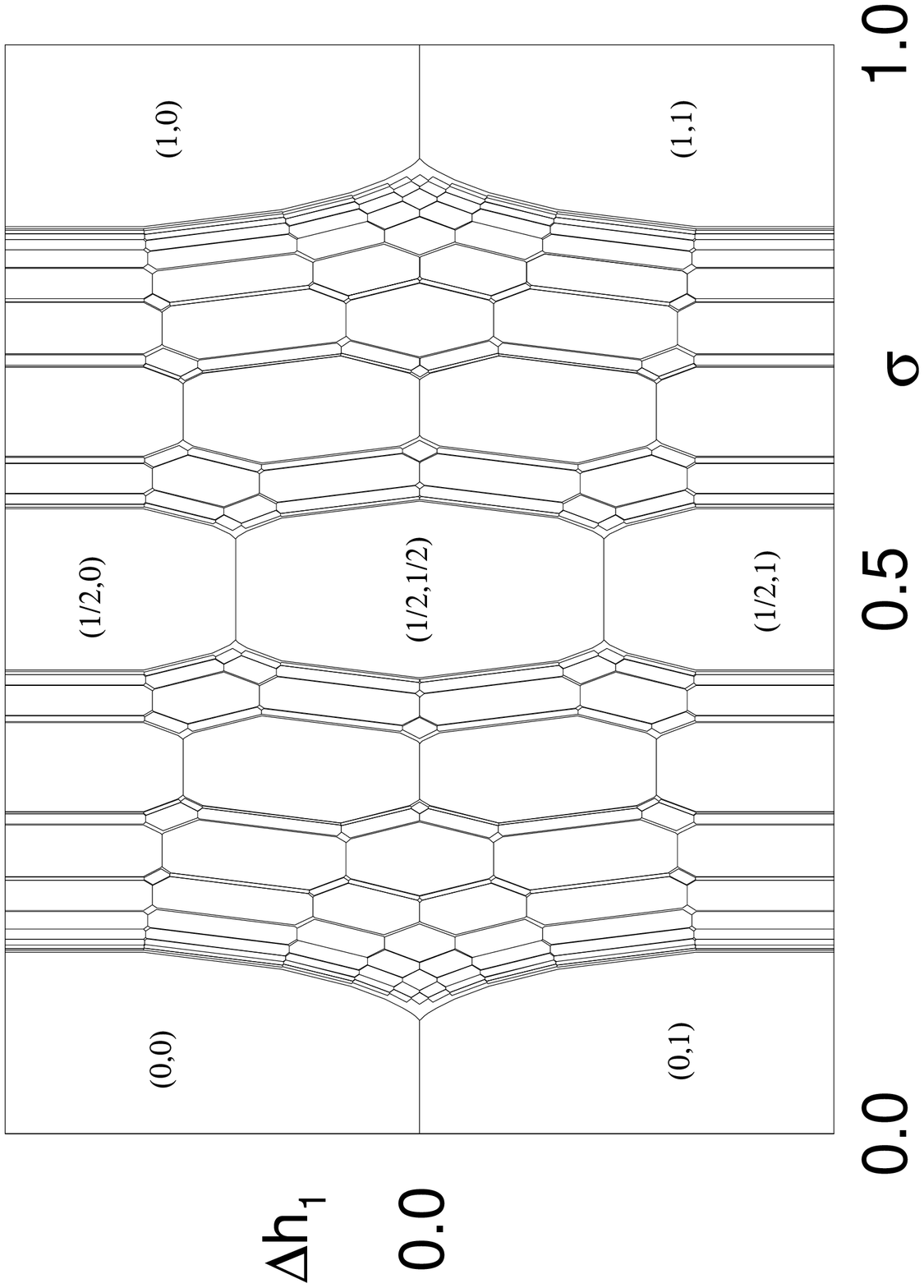}
\vskip 12.5cm
\end{figure}
\noindent
Fig.~3~~  Domains of stability (infinite-sided polygons) for 
 $(\lambda, \Delta b_1) = (0.5, 0.25)$ are shown up to $q = 10.$  The values 
of $(\omega, \nu_1)$ are inset.
\vfill\eject

\begin{figure}[h]
\includegraphics{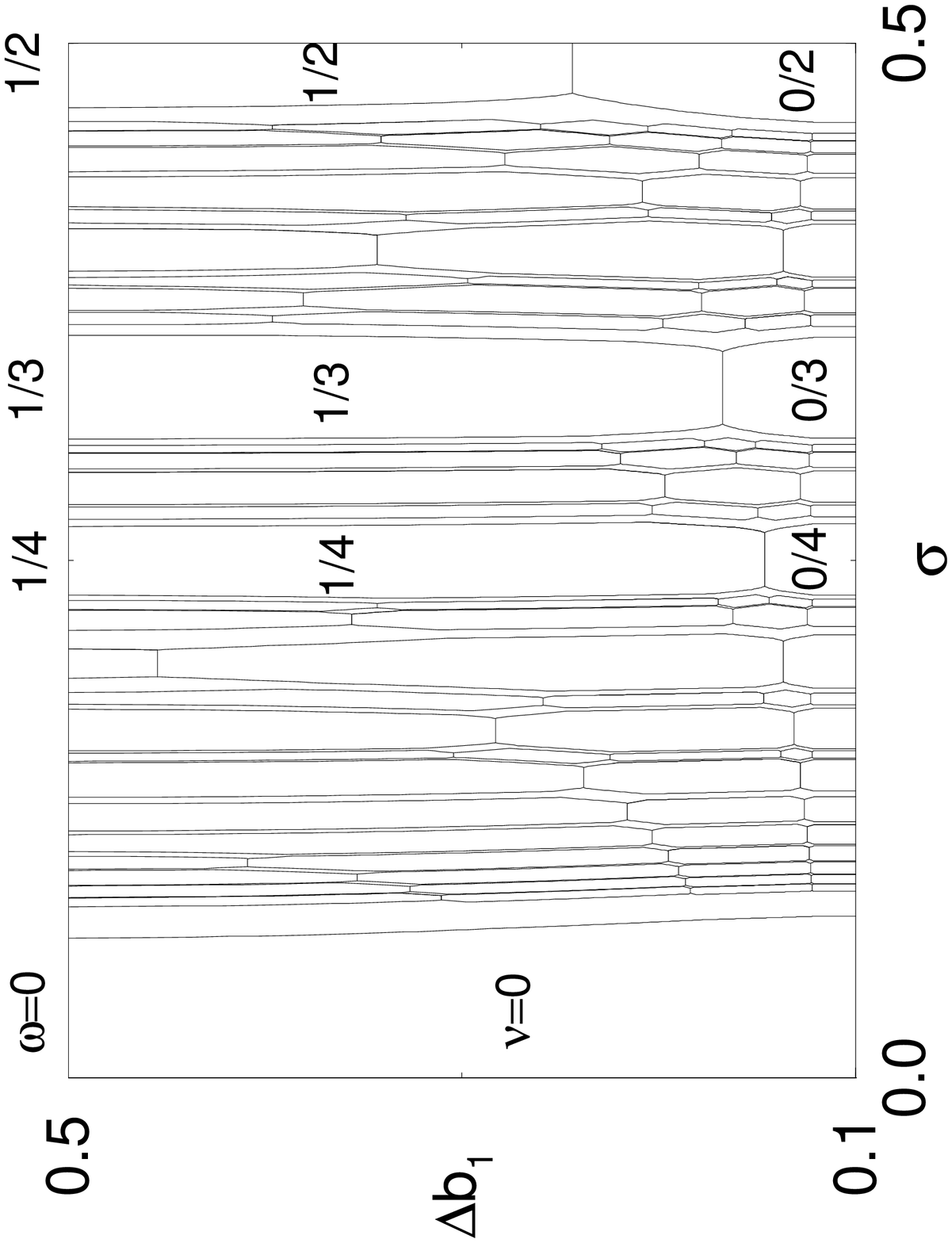}
\vskip 12.5cm
\end{figure}
\noindent
Fig.~4~~  Domains of stability for $(\lambda, \Delta h_1) = 
(0.1, 4.5 \times 10^{-2})$ are shown up to $q = 13.$  The values of $\omega$
and $\nu$ are put on the top and inset, respectively.
\vfill\eject


\begin{references}
\bibitem{fn0} Here, the relative phase is fixed by requiring
$f_\omega(0^-) \le 0 \le f_\omega(0^+).$  For detail, see\cite{KLT}.
\bibitem{fn1} Throughout the paper, we will assume $p/q$ a irreducible fraction
and the case of irrational $\omega$ can be obtained by taking $q\to \infty$.
\bibitem{fn2} Similar result for $d = 2$ is also discussed in \cite{Griffiths}.
\end{references}
\end{document}